\documentclass[letterpaper,english,aip, jmp, reprint, onecolumn]{revtex4-1}
\usepackage[T1]{fontenc}
\usepackage[latin9]{inputenc}
\usepackage{verbatim}
\usepackage{prettyref}
\usepackage{amsmath}
\usepackage{amssymb}

\makeatletter

\pdfpageheight\paperheight
\pdfpagewidth\paperwidth

\providecommand{\tabularnewline}{\\}

\usepackage{graphicx}%
\usepackage{dcolumn}%
\usepackage{bm}%

\makeatother

\usepackage{babel}
\begin{document}
\noindent\begin{minipage}[t]{1\columnwidth}%
\global\long\def\anticommutator#1#2{\left\{  #1,#2\right\}  }

\global\long\def\commutator#1#2{\left[#1,#2\right]}

\global\long\def\braket#1#2{\langle#1|#2\rangle}

\global\long\def\bra#1{\langle#1|}

\global\long\def\ket#1{|#1\rangle}

\global\long\def\Tr{\operatorname{Tr}}

\begin{comment}
Potential Changes:
\end{comment}
%
\end{minipage}

\title{Quasi-Bell inequalities from symmetrized products of noncommuting
qubit observables}

\author{Omar E. Gamel}
\email{ogamel@berkeley.edu}

\author{Graham R. Fleming}

\affiliation{Department of Chemistry, University of California, Berkeley, California
94720, USA~\\
 Molecular Biophysics and Integrated Bioimaging Division, Lawrence
Berkeley National Laboratory, Berkeley, California 94720, USA }

\date{\today}
\begin{abstract}
Noncommuting observables cannot be simultaneously measured, however,
under local hidden variable models, they must simultaneously hold
premeasurement values, implying the existence of a joint probability
distribution. We study the joint distributions of noncommuting observables
on qubits, with possible criteria of positivity and the Fr\'echet bounds
limiting the joint probabilities, concluding that the latter may be
negative. We use symmetrization, justified heuristically and then
more carefully via the Moyal characteristic function, to find the
quantum operator corresponding to the product of noncommuting observables.
This is then used to construct Quasi-Bell inequalities, Bell inequalities
containing products of noncommuting observables, on two qubits. These
inequalities place limits on local hidden variable models that define
joint probabilities for noncommuting observables. We find Quasi-Bell
inequalities have a quantum to classical violation as high as $\frac{3}{2}$,
higher than conventional Bell inequalities. The result demonstrates
the theoretical importance of noncommutativity in the nonlocality
of quantum mechanics, and provides an insightful generalization of
Bell inequalities. %
\begin{comment}
However, our approach presumes quantum theory, and therefore nonlocality,
in determining the symmetrized operator to measure. Therefore Quasi-Bell
inequalities cannot be used to experimentally test the nonlocality
of nature as traditional Bell inequalities.
\end{comment}
{} 

\begin{description}
\item [{PACS~numbers}] 03.65.Ud, 03.67.Mn, 03.67.Bg, 03.67.Lx. %
\begin{comment}
Entanglement and quantum nonlocality 03.65.Ud., Quantum Mechanics
03.65.-w, Quantum computation, 03.67.Lx. Entanglement and quantum
nonlocality in quantum information 03.67.Mn, 03.67.Bg, 
\end{comment}
{\small \par}
\end{description}
\end{abstract}


\maketitle

\section{Introduction}

Einstein, Podolsky and Rosen posed a paradox in which they showed
quantum mechanics leads to superluminal interaction, that is, nonlocality,
unless one assumes it is an incomplete theory \citep{Einstein1935}.
To preserve locality, their result suggested the existence of \emph{local
hidden variables}, inaccessible physical quantities shared between
quantum systems that were prepared together. Bell showed with his
well-known inequalities that local hidden variable models cannot reproduce
all the predictions of quantum mechanics \citep{Bell1964} demonstrating
conclusively that quantum theory is at odds with local realism. 

Clauser, Horne, Shimony, and Holt (CHSH) proposed a different Bell
inequality for a bipartite system of two spin-$\frac{1}{2}$ particles
\citep{Clauser1969} (i.e. two qubits), most commonly used today to
demonstrate quantum nonlocality theoretically and experimentally \citep{Freedman1972,Aspect1981c,Aspect1982,Aspect1982b,Hensen2015,Shalm2015b}.
It shows that a quantum expectation value violates a bound set by
local realism by a factor of $\sqrt{2}$. Further generalizations
of the CHSH inequality have been proposed \citep{Braunstein1990,Avis2005,Gisin2009},
and extended to higher Hilbert space dimensions \citep{Collins2002,Junge2009}.
But the $\sqrt{2}$ violation for the CHSH spin-$\frac{1}{2}$ case
increased only marginally \citep{Vertesi2008,Brierley2016}. 

In the construction of Bell inequalities, one assumes the independent
existence of premeasurement values. However, standard Bell inequalities
only consider products of commuting observables, that is, observables
on separate qubits. Although premeasurement values of noncommuting
observables (on the same qubit) are assumed to simultaneously exist,
they are never multiplied together. The reason for this is that the
axioms of quantum mechanics do not clearly determine the Hermitian
operator corresponding to a product of noncommuting observables. Thus
one runs into ambiguity when calculating the quantum analogue of the
classical expression in the inequality. Some may object to the existence
of such an operator, since quantum theory precludes the simultaneous
measurement of noncommuting observables. This objection fades when
we recall that\emph{ sums} of noncommuting observables clearly have
operators, without implying simultaneous measurement, and it is possible
the same is true of products. 

In this work, we explore the inclusion of products of noncommuting
observables in Bell inequalities. Such products where examined by
Arnault, although in the very different context of non-Hermitian observables
\citep{Arnault2012}. Our goal is a deeper understanding of noncommutativity
that yields insights into hidden variables, a relationship previously
addressed by Fine \citep{Fine1982a}. In quantizing the inequalities,
including the aforementioned products, we apply the symmetrization
procedure: where the quantum operator for the product of noncommuting
observables is the average of all possible permutations of the ordered
product. The implication is that the expectation value of the classical
product in a hidden variable model becomes the expectation value of
the symmetrized operator product in the quantum model. 

We justify symmetrization heuristically, and also more carefully via
Moyal quantization \citep{Moyal1949}. On application of symmetrization,
quantum theory violates our Quasi-Bell inequalities by a factor of
$\frac{3}{2}$, larger than analogous violations of traditional Bell
inequalities. The ``quasi'' prefix indicates that the inequalities
contain products of noncommuting observables, which has important
consequences in limiting experimental verification that will be discussed. 

We apply the symmetrization in the context of a spin-$\frac{1}{2}$
system with bivalent observables, i.e. where measurements yield $\pm1$,
as in \citep{Barut1988}. This is in contrast to the usual context
in which symmetrization is applied, that of the continuous conjugate
variables of position ($x$) and momentum ($p$). Before the main
results, we devote a great deal of attention to the joint probabilities
of hidden variable models of noncommuting observables on the same
spin-$\frac{1}{2}$ system. These set the context for the main results
of the paper.

The remainder of the paper is divided into two main parts. The first
part, up to and including \prettyref{sec:Symmetrization}, analyzes
local hidden variable models that include joint probabilities, which
provide the context for and build up to the symmetrization procedure.
We begin with a review of quantum and local hidden variable models
for commuting observables on two spin-$\frac{1}{2}$ particles in
\prettyref{sec:Commuting-observables}. We then shift our discussion
to noncommuting observables on one of the particles in \prettyref{sec:Non-commuting-Observables},
where we outline limits placed by positivity and the Fr\'echet bounds
on the joint probability distribution of two such observables. This
is extended to three noncommuting observables in \prettyref{sec:Three-or-more},
since this is the minimum number needed for our Quasi-Bell inequality.
These two sections take positivity and the Fr\'echet bounds as far as
they will go, showing they each lead to a trivial independent joint
probability function. This suggests we accept joint probabilities
that are potentially negative, to get a more meaningful distribution.
In this context, we introduce and justify the symmetrization procedure
in \prettyref{sec:Symmetrization}. 

The second and more important part of the paper constructs the Quasi-Bell
inequalities and finds their quantum violations. \prettyref{sec:Bell-Inequalities-Three}
reviews the CHSH inequality, and introduces a Quasi-Bell inequality
with the product of three noncommuting observables. Using results
of the Secs. \ref{sec:Three-or-more}-\ref{sec:Symmetrization}, the
quantum to classical violation factor is found to be $\frac{3}{2}$.
Higher order inequalities with more observables are constructed in
\prettyref{sec:Bell-InequalitiesN}. They are found to not increase
the violation any further. In \prettyref{sec:Classicality-of-Werner}
we discuss the relations these inequalities have to conventional Bell
inequalities and hidden variable models on Werner States, and the
implied ranges of locality and nonlocality \citep{Werner1989,Acin2006,Augusiak2014,Brunner2014}.
We conclude with a discussion of the validity, conceptual strength,
and limitations of the paper's results in \prettyref{sec:Discussion}.

We use the labels spin-$\frac{1}{2}$ particle and qubit interchangeably,
as the two are essentially equivalent. Our overall system contains
two qubits, one each for Alice and Bob. Commuting observables are
those on separate qubits, and non commuting observables are those
on the same qubit. The Pauli matrices $\sigma_{1}=\left(\begin{smallmatrix}0 & 1\\
1 & 0
\end{smallmatrix}\right)$, $\sigma_{2}=\left(\begin{smallmatrix}0 & -i\\
i & 0
\end{smallmatrix}\right)$, and $\sigma_{3}=\left(\begin{smallmatrix}1 & 0\\
0 & -1
\end{smallmatrix}\right)$, together form a vector of matrices $\vec{\sigma}$. We index multiple
noncommuting observables on the same qubit starting at 0 to facilitate
generalization of our Quasi-Bell inequalities in \prettyref{sec:Bell-InequalitiesN}. 

\section{\label{sec:Commuting-observables}Commuting observables}

We begin by reviewing both the quantum and local (classical) probability
distribution functions for two qubits. We highlight the relationship
between the two models, and the implied joint probabilities.

\subsection{Quantum probabilities}

Suppose Alice and Bob share a bipartite system of two spin-$\frac{1}{2}$
particles. They each choose the direction along which to measure spin,
$\hat{a}$, $\hat{b}$ yielding results $a,b\in\{1,-1\}$ respectively.
Their choices of measurement direction are independent of one another.
Since Alice and Bob have separate subsystems, quantum measurements
on them commute with one another.

Assume Alice and Bob share a bipartite quantum state with density
matrix $\rho$ whose Bloch matrix components are the real vectors
$\vec{u},\vec{v}$ and matrix $R$. That is, 
\[
\rho=\frac{1}{4}\left(I\otimes I+\sum_{i}u_{i}\sigma_{i}\otimes I+\sum_{j}v_{j}I\otimes\sigma_{j}+\sum_{ij}R_{ij}\sigma_{i}\otimes\sigma_{j}\right).
\]
 %
\begin{comment}
Both the quantum and local joint probability functions satisfy their
own completeness relation 
\begin{equation}
\sum_{a,b}p(a,b|\hat{a},\hat{b})=1,\label{eq:completeness}
\end{equation}
 for all $\hat{a},\hat{b}$, where $a,b$ are always summed over their
values $\pm1$. 
\end{comment}
The positivity of $\rho$ places certain conditions on $\vec{u},\vec{v},R$
\citep{Gamel2016}. The \emph{quantum} joint probability is then given
by 
\begin{equation}
p_{q}(a,b|\hat{a},\hat{b})=\Tr\left[\rho P_{a}(\hat{a})\otimes P_{b}(\hat{b})\right]=\frac{1}{4}\left(1+a\,\hat{a}\cdot\vec{u}+b\,\hat{b}\cdot\vec{v}+ab\,\hat{a}^{\dagger}R\hat{b}\right),\label{eq:quantumABdef}
\end{equation}
where $P_{a}(\hat{a})\equiv\frac{1}{2}(I+a\,\hat{a}\cdot\vec{\sigma})$
is a projection operator. 

Individual subsystem probabilities may easily be calculated from \prettyref{eq:quantumABdef}
as $p_{q}(a|\hat{a})=\frac{1}{2}\left(1+a\,\hat{a}\cdot\vec{u}\right)$,
and $p_{q}(b|\hat{b})=\frac{1}{2}\left(1+b\,\hat{b}\cdot\vec{v}\right)$.
The quantum expectation values satisfy $\langle a\rangle_{q}=\hat{a}\cdot\vec{u}$,
$\langle b\rangle_{q}=\hat{b}\cdot\vec{v}$ and $\langle ab\rangle_{q}=\hat{a}^{\dagger}R\hat{b}$.

\begin{comment}
\begin{gather}
\langle a\rangle_{q}=\sum_{a,b}a\,p_{q}(a,b|\hat{a},\hat{b})=\hat{a}\cdot\vec{u},\qquad\langle b\rangle_{q}=\sum_{a,b}b\,p_{q}(a,b|\hat{a},\hat{b})=\hat{b}\cdot\vec{v},\nonumber \\
\langle ab\rangle_{q}=\sum_{a,b}ab\,p_{q}(a,b|\hat{a},\hat{b})=\hat{a}^{\dagger}R\hat{b},\label{eq:quantumExpectation}
\end{gather}

for all $\hat{a},\hat{b}$, where $a,b$ are always summed over their
allowed values $\pm1$. 
\end{comment}

\subsection{Local hidden variables}

If local hidden variables can describe the correlation between the
two parties, then the \emph{local }joint probability is
\begin{equation}
p_{l}(a,b|\hat{a},\hat{b})=\int p(\lambda)p_{A}(a|\hat{a},\lambda)p_{B}(b|\hat{b},\lambda)d\lambda,\label{eq:localABdef}
\end{equation}
where $\lambda$ indicates the hidden variables, their distribution
$p(\lambda)$ satisfying $\int p(\lambda)d\lambda=1$, and $0\le p_{A}(a|\hat{a},\lambda),p_{B}(b|\hat{b},\lambda)\le1$
are the local probabilities of Alice and Bob respectively. 

If Alice were measuring her subsystem alone, her own local probability
would be
\begin{equation}
p_{A}(a|\hat{a})=\int p(\lambda)p_{A}(a|\hat{a},\lambda)d\lambda.\label{eq:oneSubsystem}
\end{equation}
Note that $p_{A}$ denotes two different but related probability functions,
with inclusion of the arguments removing ambiguity. %
\begin{comment}
We can also write those 
\begin{align}
p_{A}(1|\hat{a},\lambda)+p_{A}({-}1|\hat{a},\lambda) & =1,\nonumber \\
p_{A}(1|\hat{a},\lambda)-p_{A}({-}1|\hat{a},\lambda) & \equiv f_{A}(\hat{a},\lambda),\label{eq:localfdef}
\end{align}
for all $\hat{a},\lambda$, where the first line is a completeness
relation and in the second line we have defined the difference $f_{A}$,
which satisfies $-1\le f_{A}\le1$. 
\end{comment}
Without loss of generality, we can write 
\begin{equation}
p_{A}(a|\hat{a},\lambda)=\frac{1}{2}\left[1+a\,f_{A}(\hat{a},\lambda)\right],\label{eq:local_p_and_f}
\end{equation}
for function $f_{A}$ satisfying $-1\le f_{A}\le1$. The analogue
holds for Bob's local probability with notation changing accordingly. 

One can then rewrite the local joint probability \prettyref{eq:localABdef}
as 
\begin{equation}
p_{l}(a,b|\hat{a},\hat{b})=\frac{1}{4}\left(1+a\,\overline{f_{A}(\hat{a})}+b\,\overline{f_{B}(\hat{b})}+ab\,\overline{f_{A}(\hat{a})f_{B}(\hat{b})}\right),\label{eq:localABdef_f}
\end{equation}
where the overline indicates the weighted average over the hidden
variable $\lambda$, as per $\overline{f(\hat{n})}\equiv\int p(\lambda)f(\hat{n},\lambda)d\lambda$.%
\begin{comment}
Making use of \prettyref{eq:localABdef}, expectation values then
satisfy 
\begin{gather}
\langle a\rangle_{l}=\sum_{a,b}a\,p_{l}(a,b|\hat{a},\hat{b})=\overline{f_{A}(\hat{a})},\qquad\langle b\rangle_{l}=\sum_{a,b}b\,p_{l}(a,b|\hat{a},\hat{b})=\overline{f_{B}(\hat{b})},\nonumber \\
\langle ab\rangle_{l}=\sum_{a,b}ab\,p_{l}(a,b|\hat{a},\hat{b})=\overline{f_{A}(\hat{a})f_{B}(\hat{b})}.\label{eq:localExpectation}
\end{gather}
\end{comment}

We are now in a position to demand that the quantum probabilities
be simulable via a local hidden variable model. Doing so requires
equality of joint probabilities \prettyref{eq:quantumABdef} and \prettyref{eq:localABdef_f},
yielding
\begin{align}
\hat{a}\cdot\vec{u} & =\overline{f_{A}(\hat{a})}, & \hat{b}\cdot\vec{v} & =\overline{f_{B}(\hat{b})}, & \hat{a}^{\dagger}R\hat{b} & =\overline{f_{A}(\hat{a})f_{B}(\hat{b})}.\label{eq:quantum=00003Dlocal}
\end{align}
Bell's famous inequalities demonstrate that for some quantum states
(e.g. the singlet state $\vec{u}=\vec{v}=0,$ $R=-I$) hidden variables
are impossible \citep{Bell1964}. Characterizing the exact set of
$\vec{u},\vec{v},R$ for which \prettyref{eq:quantum=00003Dlocal}
has a solution for some $f_{A},f_{B}$ is equivalent to characterizing
two-qubit states simulable by local hidden variable models. Currently,
this problem only has partial solutions \citep{Augusiak2014,TonerLHV2005}.

\section{\label{sec:Non-commuting-Observables}Noncommuting Observables}

\subsection{Joint probability}

Before discussing Bell inequalities with noncommuting observables,
we must answer the following question: what can one say about joint
probabilities of noncommuting observables? More precisely, suppose
Alice makes one of two measurements along directions $\hat{a}_{0}$,
$\hat{a}_{1}$ yielding results $a_{0},a_{1}\in\{1,-1\}$. In general
these are incompatible observables due to their noncommutativity,
and cannot be measured simultaneously, despite theoretical attempts
at such a definition \citep{She1966,Prugovecki1973}. Thus the question
of a joint probability function does not arise operationally in quantum
theory. 

However, if we attempt to describe the correlations in terms of hidden
variables, that is, outcomes existing prior to measurement, then we
should be able to find a joint probability function. It is of interest
to study the properties of such a joint probability distribution if
it could exist. In fact, the existence of such a joint distribution
has been reported to be equivalent to Bell's inequalities holding
\citep{Fine1982a}. The question has also been investigated for the
continuous degrees of freedom position and momentum \citep{Ballentine1970}.
We extend this to the bivalent qubit observables at hand. 

We seek a reasonable expression for $p_{A}(a_{0},a_{1}|\hat{a}_{0},\hat{a}_{1})$,
the joint probability that Alice's measurement would yield outcomes
$a_{0},a_{1}$ respectively for measurements along directions $\hat{a}_{0}$,
$\hat{a}_{1}$. We require that the joint probability yield the correct
marginal probabilities for each measurement. For example, $\sum_{a_{1}}p_{A}(a_{0},a_{1}|\hat{a}_{0},\hat{a}_{1})=p_{A}(a_{0}|\hat{a}_{0})=\frac{1}{2}\left(1+a_{0}\,\hat{a}_{0}\cdot\vec{u}\right)$.
\begin{comment}
\begin{align}
\sum_{a_{1}}p_{A}(a_{0},a_{1}|\hat{a}_{0},\hat{a}_{1}) & =p_{A}(a_{0}|\hat{a}_{0})=\frac{1}{2}\left(1+a_{0}\,\hat{a}_{0}\cdot\vec{u}\right),\nonumber \\
\sum_{a_{0}}p_{A}(a_{0},a_{1}|\hat{a}_{0},\hat{a}_{1}) & =p_{A}(a_{1}|\hat{a}_{1})=\frac{1}{2}\left(1+a_{1}\,\hat{a}_{1}\cdot\vec{u}\right).\label{eq:noncomm_marginal}
\end{align}
\end{comment}
Making use of the known marginal probabilities and the probability
function's completeness, we can write without loss of generality
\begin{equation}
p_{A}(a_{0},a_{1}|\hat{a}_{0},\hat{a}_{1})=\frac{1}{4}\left(1+a_{0}\,\hat{a}_{0}\cdot\vec{u}+a_{1}\,\hat{a}_{1}\cdot\vec{u}+a_{0}a_{1}\,\langle a_{0}a_{1}\rangle\right).\label{eq:noncomm_form}
\end{equation}
The expectation value $\langle a_{0}a_{1}\rangle$ is some as yet
undetermined function of $\hat{a}_{0},\hat{a}_{1}$, and $\vec{u}$. 

It is instructive to compare the noncommuting joint probability function
\prettyref{eq:noncomm_form} with commuting one \prettyref{eq:quantumABdef}.
The two are very similar, with the expression for commuting observables
potentially simulating that for noncommuting ones if we set $\vec{v}=\vec{u}$
and the expectation of the product takes the form $\langle a_{0}a_{1}\rangle=\hat{a}_{0}R_{nc}\hat{a}_{1}$
for some $R_{nc}$, the ``correlation matrix'' for simulating the
noncommuting measurements.

We now turn our attention to potential criteria to determine or limit
the joint expectation value $\langle a_{0}a_{1}\rangle$. We consider
two criteria that are natural in classical probability theory; positivity,
and satisfaction of the Fr\'echet Inequalities for joint probabilities.

\begin{comment}
I can treat the noncommuting on same footing as commuting, it has
v=u, but what is R? not same positivity criteria as 2 qubit, because
you don't have ``nonlocal measurement'', or $\rho\ge0$, only weaker
(though ironically more difficult algebraically) condition that joint
probability $\ge0$

ya rabbi

symmetrization implies ``correlation matrix'' of noncommuting $R_{nc}=I$,
actually makes sense. Yet, $\vec{u}=\vec{v}$ and $R=I$ violates
all 3 positivity inequalities (2 of them always)

But of triple symmetrization ``correlation tensor'' is $W_{ijk}=\frac{1}{3}(\delta_{ij}u_{k}+\delta_{ik}u_{j}+\delta_{jk}u_{i})$,
so dependent on u. Not isotropic, but clearly that is not what matters.

what does it mean it 3 qubit vs 3 direction case?
\end{comment}
\begin{comment}
This is precisely what the Wigner function does for the incompatible
observables of position and momentum \citep{Wigner1932}, though it
yields a quasi-probability distribution that often becomes negative
(cite papers on properties of Wigner functions). What is relation
in this section with probabilities of previous section. relate the
reduced $\rho_{A}$ here with the full $\rho$ above

-Wigner function is not quite a joint probability function.

Wigner for discrete \citep{Bjork2008} 
\end{comment}
. 

\subsection{Positive distribution }

One obvious condition is the \emph{positivity} of the joint probability
distribution,
\begin{equation}
p(a_{0},a_{1}|\hat{a}_{0},\hat{a}_{1})\ge0,\label{eq:jointpositive}
\end{equation}
such that the joint probabilities are realizable non-negative values.
We enforce \prettyref{eq:jointpositive} by requiring the right hand
side of \prettyref{eq:noncomm_form} to be non-negative for the four
possible values of the pair $a_{0},a_{1}$. This yields conditions
that are instructively summarized in the following inequalities,
\begin{equation}
-(1\pm\hat{a}_{0}\cdot\vec{u})(1\pm\hat{a}_{1}\cdot\vec{u})\le\langle a_{0}a_{1}\rangle-(\hat{a}_{0}\cdot\vec{u})(\hat{a}_{1}\cdot\vec{u})\le(1\pm\hat{a}_{0}\cdot\vec{u})(1\mp\hat{a}_{1}\cdot\vec{u}),\label{eq:noncomm_positivitycondition}
\end{equation}
which hold for both the upper and lower signs. Defining the difference
quantity in the middle, 
\begin{equation}
D(\hat{a}_{0},\hat{a}_{1})\equiv\langle a_{0}a_{1}\rangle-(\hat{a}_{0}\cdot\vec{u})(\hat{a}_{1}\cdot\vec{u}),\label{eq:DifferenceDef}
\end{equation}
we seek its allowable functional forms. Since $D(\hat{a}_{0},\hat{a}_{1})$
must be basis independent, it must be a function of dot products of
$\hat{a}_{0},\hat{a}_{1}$ and $\vec{u}$. 

\paragraph{Pure state:}

In case the underlying single qubit state is pure $|\vec{u}|=1$,
the minimum of the upper bound and the maximum of the lower bound
are both zero, attained for $\hat{a}_{0}=\pm\vec{u}$ or $\hat{a}_{1}=\pm\vec{u}$.
More precisely, $D(\pm\vec{u},\hat{a}_{1})=0,\,\forall\hat{a}_{1}$
and $D(\hat{a}_{0},\pm\vec{u})=0,\,\forall\hat{a}_{0}$. It can be
shown this implies $D(\hat{a}_{0},\hat{a}_{1})$ is identically zero,
and therefore 
\begin{equation}
\langle a_{0}a_{1}\rangle\equiv(\hat{a}_{0}\cdot\vec{u})(\hat{a}_{1}\cdot\vec{u}).\label{eq:positivity-pure}
\end{equation}
That is, the expectation value of the product is equal to the product
of expectation values, meaning the two measurements are \emph{independent}.
We can use commuting measurements on two qubits to simulate these
noncommuting measurements on a single qubit if we set the correlation
matrix as an outer product $R_{nc}=\vec{u}\vec{u}^{\dagger}$, i.e.
the two qubits are in a product state. This is not surprising, since
we required positivity, and the only physical two-qubit states where
each individual qubit's state is pure are product states.

However, given that the two noncommuting measurements are on the same
qubit, this independence is highly unexpected. This is partially because
in the limit $\hat{a}_{1}\rightarrow\hat{a}_{0}$ we get $\langle a_{0}a_{1}\rangle\rightarrow(\hat{a}_{0}\cdot\vec{u})^{2}$,
which is not identically unity as $\langle a_{0}^{2}\rangle=1$ would
imply. It is intuitively expected that different measurements of the
same qubit should be at the very least correlated. If this intuition
is correct, the independence derived above casts doubt on the assumption
of positivity of the joint probability. 

In a different context, %
\begin{comment}
\citep{Ballentine1970} rejects independence of position and momentum
with an example based on scattering. 
\end{comment}
{} Ballentine found independent joint probability distributions of noncommuting
observables to satisfy positivity, but dismissed this on physical
grounds \citep{Ballentine1970}. If we follow suit and choose to reject
independence of noncommuting measurements, we will have to accept
negative probabilities, a recurring theme within quantum theory, which
we address in more detail in \prettyref{sec:Symmetrization}. 

\paragraph{Mixed state:}

In case the underlying single qubit state is mixed,$u\equiv|\vec{u}|<1$,
there is an allowed range for $D(\hat{a}_{0},\hat{a}_{1})$ and hence
$\langle a_{0}a_{1}\rangle$. In particular, \prettyref{eq:noncomm_positivitycondition}
implies

\begin{equation}
(-1+\hat{a}\cdot\vec{u})(1-u)\le D(\hat{a},\hat{u}),D(\hat{u},\hat{a})\le(1+\hat{a}\cdot\vec{u})(1-u).\label{eq:noncomm_positivitycondition_mixed}
\end{equation}
This condition will be satisfied by any $D=(\alpha+\hat{a}\cdot\vec{u})(1-u)$
for some $-1\le\alpha\le1$. For example we may define $D(\hat{a}_{0},\hat{a}_{1})\equiv\left[\hat{a}_{0}\cdot\vec{u}+\hat{a}_{1}\cdot\vec{u}+(1-u)\hat{a}_{0}\cdot\hat{a}_{1}\right](1-u)$,
where the two right most terms in the square brackets constitute $\alpha$,
and encompass the range between $1$ and $-1$. Note that this definition
of $D(\hat{a}_{0},\hat{a}_{1})$ is symmetric in its two arguments,
as one would expect. The expectation value of the product then takes
the interesting form 
\begin{equation}
\langle a_{0}a_{1}\rangle\equiv\left[\hat{a}_{0}\cdot\vec{u}+\hat{a}_{1}\cdot\vec{u}+(1-u)\hat{a}_{0}\cdot\hat{a}_{1}\right](1-u)-(\hat{a}_{0}\cdot\vec{u})(\hat{a}_{1}\cdot\vec{u}).\label{eq:positivity-mixed}
\end{equation}

If the underlying qubit is maximally mixed, $\vec{u}=0$, then \prettyref{eq:noncomm_positivitycondition}
simplifies to the weak condition $-1\le\langle a_{0}a_{1}\rangle\le1$.
In this case, there is a great deal of freedom in assigning a functional
form to $\langle a_{0}a_{1}\rangle$ that always lies within this
range. We may follow the form of \prettyref{eq:positivity-mixed},
which yields $\langle a_{0}a_{1}\rangle\equiv\hat{a}_{0}\cdot\hat{a}_{1}$
for $u=0$, and incidentally corresponds to symmetrization of noncommuting
observables, the subject of \prettyref{sec:Symmetrization}. 

One can conclude that if the quantum state is mixed, requiring the
joint probability of noncommuting observables to be positive does
not necessarily imply independence. This is interesting in its own
right, and opens the door for reasonable and positive joint probabilities
for some quantum states. However, the implied independence for pure
states means one cannot demand positivity in general.

\begin{comment}
This \textquotedbl{}operators either commute or they don't\textquotedbl{}
paradigm is rather strange. Shouldn't commutativity be a spectrum,
and this should have a bearing on their simultaneous measurement (it
already does on uncertainty principle, since Heisenberg's formula
includes the magnitude of the commutator) 
\end{comment}

\subsection{Fr\'echet Inequalities}

Possible values for the joint probability of two classical events
are bound by the individual (marginal) probability of each event.
For example, the joint probabilities of two events each with probability
unity (zero) must itself be unity (zero). The joint probability of
two events each with probability $\frac{1}{2}$ may be $0$ if they
are mutually exclusive, $\frac{1}{4}$ if they are independent, $\frac{1}{2}$
if they fully coincide, or any value in between.%
\begin{comment}
Mentioned by \citep{Janssens2004} but generically, not analyzed for
qubit observables
\end{comment}

More precisely, classical joint probabilities must satisfy the Fr\'echet
inequalities \citep{Frechet1935,Frechet1951}, which place bounds
based on the individual probabilities:

\begin{equation}
p(a_{0}|\hat{a}_{0})+p(a_{1}|\hat{a}_{1})-1\le p(a_{0},a_{1}|\hat{a}_{0},\hat{a}_{1})\le\min\{p(a_{0}|\hat{a}_{0}),p(a_{1}|\hat{a}_{1})\}.\label{eq:Frechet}
\end{equation}
Plugging the marginal and joint probabilities implied by \prettyref{eq:noncomm_form}
into \prettyref{eq:Frechet}, rearranging and simplifying yields the
following two inequalities
\begin{gather}
\frac{1}{4}\left(1-a_{0}\,\hat{a}_{0}\cdot\vec{u}-a_{1}\,\hat{a}_{1}\cdot\vec{u}+a_{0}a_{1}\,\langle a_{0}a_{1}\rangle\right)\ge0,\label{eq:Frechetlower}\\
\frac{1}{4}\left(1+\min\{a_{0}\,\hat{a}_{0}\cdot\vec{u},a_{1}\,\hat{a}_{1}\cdot\vec{u}\}-\max\{a_{0}\,\hat{a}_{0}\cdot\vec{u},a_{1}\,\hat{a}_{1}\cdot\vec{u}\}-a_{0}a_{1}\,\langle a_{0}a_{1}\rangle\right)\ge0.\label{eq:FrechetUpper}
\end{gather}

Upon comparing with \prettyref{eq:noncomm_form}, it is evident the
left hand side of \prettyref{eq:Frechetlower} is equal to $p(-a_{0},-a_{1}|\hat{a}_{0},\hat{a}_{1})$.
Similarly, the left hand side of \prettyref{eq:FrechetUpper} is equal
to $p(-a_{0},a_{1}|\hat{a}_{0},\hat{a}_{1})$ or $p(a_{0},-a_{1}|\hat{a}_{0},\hat{a}_{1})$
depending whether $a_{0}\,\hat{a}_{0}\cdot\vec{u}$ or $a_{1}\,\hat{a}_{1}\cdot\vec{u}$
is larger. Since all the above inequalities are meant to hold $\forall a_{0},a_{1}\in\{1,-1\}$,
\prettyref{eq:Frechetlower} and \prettyref{eq:FrechetUpper} are
each equivalent to \prettyref{eq:jointpositive}. In other words,\emph{
requiring the Fr\'echet inequalities hold is identical to the requirement
of positivity of the joint probability}, which as we showed in the
previous section, has the unwanted consequence of independence for
pure states. 

\section{\label{sec:Three-or-more}Three or more observables }

Suppose we increase the number of noncommuting observables, starting
with three. A derivation similar to that of \prettyref{eq:noncomm_form}
will show that the triple joint probability, without loss of generality,
can be written as 
\begin{equation}
p(a_{0},a_{1},a_{2}|\hat{a}_{0},\hat{a}_{1},\hat{a}_{2})=\frac{1}{8}\big(1+a_{0}\,\hat{a}_{0}\cdot\vec{u}+a_{1}\,\hat{a}_{1}\cdot\vec{u}+a_{2}\,\hat{a}_{2}\cdot\vec{u}+a_{0}a_{1}\,\langle a_{0}a_{1}\rangle+a_{0}a_{2}\,\langle a_{0}a_{2}\rangle+a_{1}a_{2}\,\langle a_{1}a_{2}\rangle+a_{0}a_{1}a_{2}\,\langle a_{0}a_{1}a_{2}\rangle\big).\label{eq:noncomm_form-three}
\end{equation}
If we require this joint probability to always be nonnegative, we
may engage in a derivation similar to that of inequality \prettyref{eq:noncomm_positivitycondition}.
For pure states $|\vec{u}|=1$, the result will be analogous to the
two-observable case. That is
\begin{equation}
\langle a_{0}a_{1}a_{2}\rangle\equiv(\hat{a}_{0}\cdot\vec{u})(\hat{a}_{1}\cdot\vec{u})(\hat{a}_{2}\cdot\vec{u}).\label{eq:independent-threeq}
\end{equation}
Thus, requiring positivity of the joint probability distribution of
three noncommuting observables for a pure state also implies their
independence. The same procedure can be extended to show the independence
of any number of noncommuting observables on a pure state, if positivity
is required.

However, it is interesting that the two Fr\'echet inequalities when
applied to three (or more) noncommuting observables are not both equal
to the positivity condition, as was the case for two observables.
Given three observables, the two-observable Fr\'echet inequalities \prettyref{eq:Frechet}
will still hold for all pairs. The additional three-observable Fr\'echet
inequalities are
\begin{equation}
\max_{ijk}\{p(a_{i}|\hat{a}_{i})+p(a_{j},a_{k}|\hat{a}_{j},\hat{a}_{k})-1\}\le p(a_{0},a_{1},a_{2}|\hat{a}_{0},\hat{a}_{1},\hat{a}_{2})\le\min_{mn}\{p(a_{m},a_{n}|\hat{a}_{m},\hat{a}_{n})\},\label{eq:Frechet-Three}
\end{equation}
where the indices $i,j,k$ are distinct, $l,m,n$ are distinct, and
all take values $0,1,2$. Note that \prettyref{eq:Frechet-Three}
is obtained from \prettyref{eq:Frechet} by treating occurrence $(a_{0},a_{1},a_{2}|\hat{a}_{0},\hat{a}_{1},\hat{a}_{2})$
as a two-way conjunction of $(a_{i}|\hat{a}_{i})$ and $(a_{j},a_{k}|\hat{a}_{j},\hat{a}_{k})$.
We could also treat it as a three-way conjunction of $(a_{i}|\hat{a}_{i})$,
$(a_{j}|\hat{a}_{j})$, and $(a_{k}|\hat{a}_{k})$. However, this
more reductionist approach yields a weaker inequality that is implied
by \prettyref{eq:Frechet-Three} and \prettyref{eq:Frechet} anyway. 

We proceed by plugging \prettyref{eq:noncomm_form-three} and \prettyref{eq:noncomm_form}
into \prettyref{eq:Frechet-Three}, rearranging and simplifying. This
yields the following two inequalities for the lower and upper bound
respectively,
\begin{gather}
\frac{1}{8}\big(3-3a_{i}\,\hat{a}_{i}\cdot\vec{u}-a_{j}\,\hat{a}_{j}\cdot\vec{u}-a_{k}\,\hat{a}_{k}\cdot\vec{u}+a_{i}a_{j}\,\langle a_{i}a_{j}\rangle+a_{i}a_{k}\,\langle a_{i}a_{k}\rangle-a_{j}a_{k}\,\langle a_{j}a_{k}\rangle+a_{i}a_{j}a_{k}\,\langle a_{i}a_{j}a_{k}\rangle\big)\ge0,\label{eq:Frechetlower-three}\\
\frac{1}{8}\big(1-a_{l}\,\hat{a}_{l}\cdot\vec{u}+a_{m}\,\hat{a}_{m}\cdot\vec{u}+a_{n}\,\hat{a}_{n}\cdot\vec{u}-a_{l}a_{m}\,\langle a_{l}a_{m}\rangle-a_{l}a_{n}\,\langle a_{l}a_{n}\rangle+a_{m}a_{n}\,\langle a_{m}a_{n}\rangle-a_{l}a_{m}a_{n}\,\langle a_{l}a_{m}a_{n}\rangle\big)\ge0,\label{eq:FrechetUpper-three}
\end{gather}
where the indices $i,j,k$ and $l,m,n$ are assumed to be the ones
satisfying the maximum / minimum in \prettyref{eq:Frechet-Three}.
These two bounds can be rewritten as
\begin{gather}
p(-a_{i},-a_{j},-a_{k}|\hat{a}_{i},\hat{a}_{j},\hat{a}_{k})+p(-a_{i},-a_{j},a_{k}|\hat{a}_{i},\hat{a}_{j},\hat{a}_{k})+p(-a_{i},a_{j},-a_{k}|\hat{a}_{i},\hat{a}_{j},\hat{a}_{k})\ge0,\label{eq:Frechetlower-threePos}\\
p(-a_{l},a_{m},a_{n}|\hat{a}_{l},\hat{a}_{m},\hat{a}_{n})\ge0.\label{eq:FrechetUpper-threePos}
\end{gather}
The lower Fr\'echet bound \prettyref{eq:Frechetlower-threePos} is implied
by, but weaker than the positivity condition. The upper bound \prettyref{eq:FrechetUpper-threePos}
is equivalent to the positivity condition. 

Therefore, we may conclude that for three observables, the Fr\'echet
bounds taken together, are again equivalent to the positivity condition.
This seems to hold for more observables as well.

\begin{comment}
Related to Fr\'echet-Hoeffding inequalities \citep{Hoeffding1940,Hoeffding1941,Frechet1951}

But this still means Frechet imply positivity.
\end{comment}

\section{\label{sec:Symmetrization}Symmetrization}

Thus far, we considered requiring the joint probabilities to be non-negative
or, equivalently, that they satisfy the Fr\'echet inequalities of classical
probability theory. It was found that in the case of pure quantum
states, these imply the independence of measurements of any noncommuting
observables. We deemed this independence unsatisfactory on physical
grounds, and now seek a different approach. It is obvious that any
approach that does not explicitly require positivity of the joint
distribution may yield some negative probabilities. Though physically
meaningless, negative probabilities need not be an operational problem
since they correspond to impossible simultaneous measurements \citep{Al-Safi2013}.
In Feynman's words, they may be used if ``the situation for which
the probability appears to be negative is not one that can be verified
directly'' \citep{Feynman1987}.

The negativity of joint probabilities of noncommuting observables
is well known \citep{Broglie1960,Margenau1961,Prugovecki1973,Khrennikov1997,Levy2015}.
It has even been shown as equivalent to fundamental nonclassical features
of quantum theory \citep{K.Wodkiewicz1988b,Sudarshan1993,Rothman2001,Spekkens2008,Veitch2012}.
This phenomenon is also related to the Wigner quasi-probability distribution
taking negative values \citep{Wigner1932,Hudson1974}. 

As per \prettyref{eq:noncomm_form}. Thus we need an expression for
$\langle a_{0}a_{1}\rangle$ based on the quantum expectation value.
At first glance, the product of noncommuting observables which cannot
be simultaneously measured may seem precluded by quantum mechanics.
However, consider that the sum of noncommuting observables is a well-defined
observable in its own right, e.g. $\sigma_{1}+\sigma_{2}$. As Bell
argued, ``a measurement of a sum of noncommuting observables cannot
be made by combining trivially the results of separate observations
on the two terms \textemdash{} it requires a quite distinct experiment''
\citep{Bell1966}. Similarly, the product does not involve measuring
each operator and multiplying the results, rather the product is itself
a legitimate observable whose operator can be derived in a manner
consistent with quantization rules. The question is then finding such
an operator. 

Some authors have explored joint quasiprobability distributions for
spin-$\frac{1}{2}$ states \citep{Chandler1992b,Scully1994,TerraCunha2001}.
As expected, the quasiprobabilties may be negative. However, some
of these methods lack symmetry in the arguments, or basis independence.
Others do not easily lend themselves to variable directions of the
spin operators. Therefore we pursue a more flexible approach.

\subsection{\label{subsec:Heuristic-Derivation}Heuristic Derivation}

\begin{comment}
Even though the\emph{ axioms of quantum theory} do not provide an
unambiguous clear , 
\end{comment}
Commonly used heuristic arguments consistent with quantum theory provide
\emph{symmetrization} as the quantization of the product of two noncommuting
observables, such as position and momentum. That is, the quantization
of the product is set equal to the average of all possible permutations
of the product of the quantizations. The result is Hermitian, symmetric
in all the operators, and reduces to the simple product if the operators
commute. For example $xp\rightarrow\frac{1}{2}(\hat{x}\hat{p}+\hat{p}\hat{x})$.
Many authors have made use of such a symmetrization \citep{Shewell1959,Margenau1967,Barut1988,Gustafson2011,Schwindt2016}.
Some ambiguity arises if any of the quantities in the product are
raised to a power greater than unity. However, this is not a concern
here. 

Note there are two related but subtly different concepts here, both
involving a classical quantity in some way underlying a quantum operator.
One is when a classical expression is \emph{quantized }into a quantum
observable, and the other is when a classical hidden variable is posited
to model the measurement results on said observable. It is in the
former case that symmetrization is traditionally used, while here
we use it in the latter, as did others \citep{Margenau1961,Barut1988}. 

\begin{comment}
Suppose we have three classical quantities quantized as $a,b,c\rightarrow A,B,C$.
The quantized operator of the product $abc$ must be symmetric in
$a,b,c$ to be well defined. As such, consider the symmetric identity:
\begin{equation}
24abc\equiv(a+b+c)^{3}+(a-b-c)^{3}+(-a+b-c)^{3}+(-a-b+c)^{3}.\label{eq:identityCube}
\end{equation}

We then quantize \prettyref{eq:identityCube}, applying oft-used assumptions
that quantization (i) is linear (e.g. $a-b\rightarrow A-B$), and
(ii) commutes with the integer power (e.g. $a^{n}\rightarrow A^{n}$).
The result is to capitalize the symbols on the right side of \prettyref{eq:identityCube}.
Simplifying, the result of the quantization of the product is the
uniquely defined \emph{fully symmetrized product} of the quantizations
\begin{equation}
abc\rightarrow\frac{1}{6}\left(ABC{+}BCA{+}CAB{+}CBA{+}ACB{+}BAC\right).\label{eq:cubes}
\end{equation}

\textendash but the above is problematic, because it is not unique.
specifically
\[
16abc=[(a+b)^{2}-(a-b)^{2}+c]^{2}-[(a+b)^{2}-(a-b)^{2}-c]^{2}
\]
upon this quantization would imply 
\[
abc\rightarrow\frac{1}{4}\anticommutator{\anticommutator AB}C=\frac{1}{6}\left(ABC{+}CAB{+}CBA{+}BAC\right).
\]

No $C$ in middle terms. This can be countered by demanding symmetry,
but you can demand that and get the products the first place without
these sum powers.
\end{comment}

Going forward, we now find the quantum operator that corresponds to
the classical product $a_{0}a_{1}$, with the intention of applying
standard quantum measurements to get the joint probability function.
The quantizations of $a_{0},a_{1}$ are $\hat{a}_{0}\cdot\vec{\sigma},\:\hat{a}_{1}\cdot\vec{\sigma}$
respectively. Applying symmetrization for two observables, the result
is simply half the anticommutator, yielding,
\begin{equation}
a_{0}a_{1}\rightarrow\frac{1}{2}\anticommutator{\hat{a}_{0}\cdot\vec{\sigma}}{\hat{a}_{1}\cdot\vec{\sigma}}=\hat{a}_{0}\cdot\hat{a}_{1}I.\label{eq:symmetrization-two}
\end{equation}
The symmetrized observable is proportional to the identity matrix,
with two identical eigenvalues $\hat{a}_{0}\cdot\hat{a}_{1}$, not
$\pm1$ as the individual realizations of the classical hidden variable
product $a_{0}a_{1}$. Nonetheless, there is no inconsistency. To
see this, consider sums instead of products once more. The fact that
the eigenvalues of $\sigma_{1}+\sigma_{2}$ are not sums of eigenvalues
of $\sigma_{1}$ and $\sigma_{2}$ does not preclude a local hidden
variable model, whose individual realizations yield eigenvalues of
$\sigma_{1}$ and $\sigma_{2}$, but not of their sum. 

This is the essence of Bell's refutation of von Neumann's purported
proof for the impossibility of hidden variables \citep{VonNeumann1932}.
Bell argued there is no reason to require the individual realizations
of a hidden variable model to be additive over sums of noncommuting
observables, since the observables cannot be measured simultaneously
anyway. The function of a hidden variable is model ``to reproduce
the \emph{measurable }peculiarities of quantum mechanics when \emph{averaged
over}'' \citep{Bell1966}, that is, to reproduce the expectation
values. Similar reasoning may be applied to products of noncommuting
observables. 

To be fully consistent with the statistical predictions of quantum
mechanics, $a_{0}$, $a_{1}$ of the hidden variable model must be
correlated with one another in such a way that classical expectation
values of their product coincide with the quantum expectation values
of its quantization, which as in \prettyref{eq:symmetrization-two},
is the symmetrized product operator. Aside however, note that the
analogy of the sum to product breaks down in that the expectation
value of sum is the sum of expectation values, but the expectation
value of the product is not related to the individual expectation
values in a simple way.

Moving forward, we find the expectation value of the product is then

\begin{equation}
\langle a_{0}a_{1}\rangle=\Tr\left[\rho_{A}\hat{a}_{0}\cdot\hat{a}_{1}I\right]=\hat{a}_{0}\cdot\hat{a}_{1},\label{eq:symmetrization-two-exp}
\end{equation}
which very interestingly, is independent of Alice's reduced (single
qubit) quantum state $\rho_{A}\equiv\Tr_{B}[\rho]$, and its Bloch
vector $\vec{u}$. 

The joint probability associated with symmetrization is then

\begin{eqnarray}
p(a_{0},a_{1}|\hat{a}_{0},\hat{a}_{1}) & = & \Tr\left[\rho_{A}\frac{1}{2}\anticommutator{P_{a_{0}}(\hat{a}_{0})}{P_{a_{1}}(\hat{a}_{1})}\right]\nonumber \\
 & = & \frac{1}{4}\left(1+a_{0}\,\hat{a}_{0}\cdot\vec{u}+a_{1}\,\hat{a}_{1}\cdot\vec{u}+a_{0}a_{1}\,\hat{a}_{0}\cdot\hat{a}_{1}\right).\label{eq:symmetrization-two-jointp}
\end{eqnarray}
As expected, the joint probability in \prettyref{eq:symmetrization-two-jointp}
is sometimes negative. For example if $a_{0}=a_{1}=-1$, $\hat{a}_{0}=(1,0,0)$,
$\hat{a}_{1}=(0,1,0)$, $\vec{u}=(1,1,0)/\sqrt{2},$ then it yields
$p(a_{0},a_{1}|\hat{a}_{0},\hat{a}_{1})=(1-\sqrt{2})/4=-0.104$. We
reiterate that the negative probability is unobservable directly since
the qubit cannot be measured along $\hat{a}_{0}$ and $\hat{a}_{1}$
simultaneously, and hence does not lead to operational contradictions. 

Now turning our attention to three noncommuting observables, we seek
the quantum operator for the classical product $a_{0}a_{1}a_{2}$.
Let $S$ be the set of the six possible permutations of $0,1,2$.
Then symmetrization yields%
\begin{comment}
\begin{eqnarray*}
 &  & \frac{1}{6}\sum_{lmn\in S}\big(\hat{a}_{l}\cdot\vec{\sigma}\big)\big(\hat{a}_{m}\cdot\vec{\sigma}\big)\big(\hat{a}_{n}\cdot\vec{\sigma}\big)\\
 & = & \frac{1}{6}\sum_{lmn\in S}a_{li}a_{mj}a_{nk}\sigma_{i}\sigma_{j}\sigma_{k}\\
 & = & \frac{1}{6}\sum_{lmn\in S}a_{li}a_{mj}a_{nk}\big(\delta_{ij}\sigma_{k}-\delta_{ik}\sigma_{j}+\delta_{jk}\sigma_{i}+i\varepsilon_{ijk}I\big)\\
 & = & \frac{1}{6}\sum_{lmn\in S}\left[\big(\hat{a}_{l}\cdot\hat{a}_{m}\big)\hat{a}_{n}-\big(\hat{a}_{l}\cdot\hat{a}_{n}\big)\hat{a}_{m}+\big(\hat{a}_{m}\cdot\hat{a}_{n}\big)\hat{a}_{l}\right]\cdot\vec{\sigma}\\
 & = & \frac{1}{6}\sum_{lmn\in S}\left[\big(\hat{a}_{l}\cdot\hat{a}_{m}\big)\hat{a}_{n}\right]\cdot\vec{\sigma}\\
 & = & \frac{1}{3}\left[(\hat{a}_{1}\cdot\hat{a}_{2})\hat{a}_{0}+(\hat{a}_{1}\cdot\hat{a}_{2})\hat{a}_{0}+(\hat{a}_{1}\cdot\hat{a}_{2})\hat{a}_{0}\right]\cdot\vec{\sigma}\\
 & \equiv & \vec{a}_{012}\cdot\vec{\sigma},
\end{eqnarray*}

We make use of Einstein summation notation over repeated indices $i,\,j,\,k,$
summing from $1$ to $3$.
\end{comment}

\begin{equation}
a_{0}a_{1}a_{2}\rightarrow\frac{1}{6}\sum_{lmn\in S}\big(\hat{a}_{l}\cdot\vec{\sigma}\big)\big(\hat{a}_{m}\cdot\vec{\sigma}\big)\big(\hat{a}_{n}\cdot\vec{\sigma}\big)=\vec{a}_{012}\cdot\vec{\sigma},\label{eq:symmetrization-three}
\end{equation}
where the symmetric product vector $\vec{a}_{012}$ is defined as
\begin{equation}
\vec{a}_{012}\equiv\frac{1}{3}\left[(\hat{a}_{1}{\cdot}\hat{a}_{2})\hat{a}_{0}+(\hat{a}_{2}{\cdot}\hat{a}_{0})\hat{a}_{1}+(\hat{a}_{0}{\cdot}\hat{a}_{1})\hat{a}_{2}\right],\label{eq:a012Def}
\end{equation}
fully symmetric in $\hat{a}_{0}$, $\hat{a}_{1}$ and $\hat{a}_{2}$
as required. Note that $|\vec{a}_{012}|\le1$, and the eigenvalues
of $\vec{a}_{012}{\cdot}\vec{\sigma}$ are $\pm|\vec{a}_{012}|$.
This triple symmetrization has previously been applied by Barut et
al. \citep{Barut1988}. 

The expectation value of the triple product is then

\begin{equation}
\langle a_{0}a_{1}a_{2}\rangle=\Tr\left[\rho_{A}\vec{a}_{012}\cdot\vec{\sigma}\right]=\vec{a}_{012}\cdot\vec{u}.\label{eq:symmetrization-three-exp}
\end{equation}
Unlike the product of two noncommuting observable in \prettyref{eq:symmetrization-two-exp},
the triple product in \prettyref{eq:symmetrization-three-exp} depends
on the Bloch vector $\vec{u}$, more in line with intuitive expectation.

Based on the commutation properties of the Pauli matrices, one can
generalize these properties of product of noncommuting observables
$a_{0}a_{1}a_{2}\ldots a_{N}$. If the number of operators $N+1$
is even, the symmetrization is always an operator proportional to
the identity matrix, resulting in an expectation value independent
of $\vec{u}$. If $N+1$ is odd, the symmetrization can be written
as an operator $\vec{a}_{01\ldots N}\cdot\vec{\sigma}$, with an expectation
value $\vec{a}_{01\ldots N}\cdot\vec{u}$. It is the odd case that
is relevant for the Quasi-Bell inequalities.

\begin{comment}
$\vec{a}_{01\ldots N}$ is an average of $(N+1)!!$ terms

(related to, though not equivalent to contextuality). 

One may object the full symmetrization's eigenvalues are not products
of allowable values (eigenvalues) of each. This objection applies
not only products, but even sums, e.g. . This was a critical element
of von Neumann's purported proof for the impossibility of hidden variables
\citep{VonNeumann1932}. Despite being accepted for over three decades,
it was shown to be invalid by John Bell \citep{Bell1966}. Bell reasoned
that while indeed no LHV model can replicate the non-additivity of
observable allowable values (eigenvalues), there is no reason to demand
this from an LHV model in the first place. The function of a hidden
variable is model ``to reproduce the \emph{measurable }peculiarities
of quantum mechanics when \emph{averaged over}''. 
\end{comment}

\subsection{\label{subsec:MoyalCharFunction}Moyal Characteristic Function Derivation}

The heuristic application of symmetrization, though substantiated,
may be objected to as containing an element of speculation. %
\begin{comment}
However, this is unavoidable when exploring new avenues in the fundamental
nature of quantum mechanics. The constructs we have introduced yield
interesting insights into the relationship of quantum noncommutativity
and nonlocality.
\end{comment}
{} We therefore produce a more concrete derivation, based on Moyal's
seminal representation of quantum mechanics as a statistical theory
\citep{Moyal1949}. Our derivation is simpler and more flexible than
that of Chandler et al. \citep{Chandler1992b}, which applied Moyal's
full Fourier approach to the limited case of mutually orthogonal spin
directions.

We begin by defining the \emph{characteristic function} of the three
classical quantities $a_{0}$, $a_{1}$, $a_{2}$, given by
\begin{equation}
M(\theta_{0},\theta_{1},\theta_{2})\equiv\langle e^{i(\theta_{0}a_{0}+\theta_{1}a_{1}+\theta_{2}a_{2})}\rangle,\label{eq:charFuncClassic}
\end{equation}
where the classical expectation value is over the $a_{i}$, and $\theta_{i}$
are real parameters. 

The characteristic function is a standard construct in statistics
\citep{Lukacs1972}, useful for calculation of moments. Of interest
to us, the first joint moment of $a_{0}$, $a_{1}$, $a_{2}$ is given
by the mixed partial derivative of the characteristic function evaluated
at zero, as per
\begin{equation}
\langle a_{0}a_{1}a_{2}\rangle=\left[\frac{\partial^{3}}{i^{3}\partial\theta_{0}\partial\theta_{1}\partial\theta_{2}}M(\theta_{0},\theta_{1},\theta_{2})\right]^{\theta_{i}{=}0},\label{eq:MomentClassic}
\end{equation}
where $\theta_{i}{=}0$ denotes the evaluation $\theta_{0}{=}\theta_{1}{=}\theta_{2}{=}0$. 

One can then apply Moyal's novel idea of quantizing the characteristic
function to the problem at hand. We can unambiguously quantize \prettyref{eq:charFuncClassic}
in the standard manner of replacing the classical quantities with
quantum observables. In doing so, we end up with the quantum characteristic
function
\begin{equation}
M_{\psi}(\theta_{0},\theta_{1},\theta_{2})\equiv\langle\psi|\,e^{i(\theta_{0}\hat{a}_{0}+\theta_{1}\hat{a}_{1}+\theta_{2}\hat{a}_{2})\cdot\vec{\sigma}}\,|\psi\rangle,\label{eq:charFuncQuantum}
\end{equation}
where $|\psi\rangle$ is the quantum state. Replacing $M$ in \prettyref{eq:MomentClassic}
with $M_{\psi}$, we can then calculate first joint moment of the
quantum characteristic function to get the \emph{quantum expectation
value of the product} , as
\begin{equation}
\langle a_{0}a_{1}a_{2}\rangle_{\psi}=\langle\psi|\left[\frac{\partial^{3}}{i^{3}\partial\theta_{0}\partial\theta_{1}\partial\theta_{2}}e^{i(\theta_{0}\hat{a}_{0}+\theta_{1}\hat{a}_{1}+\theta_{2}\hat{a}_{2})\cdot\vec{\sigma}}\right]^{\theta_{i}{=}0}|\psi\rangle.\label{eq:MomentQuantum}
\end{equation}

It is clear from \prettyref{eq:MomentQuantum} that the quantum operator
corresponding to the product $a_{0}a_{1}a_{2}$ is simply the one
whose expectation value is calculated on the right hand side. Relegating
the algebra to the Appendix, this operator reduces to precisely $\vec{a}_{012}\cdot\vec{\sigma}$
defined in \prettyref{eq:symmetrization-three}. Therefore, the Moyal
quantization yields exactly the same result as heuristic symmetrization.
Although this equivalence is remarkable in many ways, it may be seen
as following from the characteristic function's symmetry in $a_{0}$,
$a_{1}$, $a_{2}$ and the subsequent application of canonical quantization. 

A derivation analogous to the above can be performed for the product
of any number of spin observables. The result is identical to heuristic
symmetrization in each case.

We emphasize that our derivation above only made use of existing concepts
in classical probability theory and standard canonical quantization.
Therefore symmetrization is not an additional concept per se, rather
it arises from the combination of old concepts in a new way, and logically
follows from established quantum theory. 

That said, canonical quantization for all its successes, is known
to create ambiguity in some cases \citep{Groenewold1946,Gotay1996}.
It is uncertain whether Moyal's and our choice of applying quantization
to the exponential in the characteristic function is consistent with
applying it to other possible functions. Therefore, although symmetrization
is straightforward mathematically, this does not necessarily mean
it corresponds to physical reality. 

It may be possible to test symmetrization experimentally, by evolving
real systems whose Hamiltonians contain products of noncommuting observables,
and comparing the experimental results with theoretical simulations
based on symmetrization. If the two approaches agree, this would be
the ideal justification of symmetrization. However, it is not clear
how one can create such Hamiltonians, making this a challenging and
interesting potential research project.

\begin{comment}
Perhaps Moyal expression only works when when the $\theta_{i}$ somehow
independent, or perpendicular?

actually Moyal is right, then this can actually tell us something
about evaluating this crazy integral up there (Moyal characteristic
function just tells you where to take the integral, i.e. write as
exponential, take partials)
\end{comment}

\section{\label{sec:Bell-Inequalities-Three}Quasi-Bell inequalities: three
observables}

We are now in a position to construct the main result of this paper,
the Quasi-Bell inequalities involving products of noncommuting observables.
They yield a quantum to classical violation higher than analogous
Bell inequalities that don't use noncommuting products. 

We start with the traditional CHSH inequality. Consider classical
quantities $a_{0},a_{1},b_{0},b_{1}\in\{1,-1\}$, two each for Alice
and Bob. Define $L=a_{0}b_{0}+a_{0}b_{1}+a_{1}b_{0}-a_{1}b_{1}$.
It can be factored to
\begin{equation}
L=a_{0}(b_{0}+b_{1})+a_{1}(b_{0}-b_{1}).\label{eq:Lfactored}
\end{equation}
Of the two bracketed terms in \prettyref{eq:Lfactored}, one must
be $0$ and the other $\pm2$. Then $L=\pm2$ for any $a_{0},a_{1},b_{0},b_{1}$.

Extending to the classical probabilistic case, suppose that $a_{0},a_{1},b_{0},b_{1}$
each have some probability of being $1$ or $-1$. In other words,
there exists an ensemble of possible realizations, each element of
which has $L=\pm2$. Then the \emph{classical expectation value} over
the whole ensemble must satisfy $-2\le\langle L\rangle\le2$. More
precisely 
\begin{equation}
|\langle L\rangle|=|\langle a_{0}b_{0}\rangle+\langle a_{0}b_{1}\rangle+\langle a_{1}b_{0}\rangle-\langle a_{1}b_{1}\rangle|\le2,\label{eq:CHSH}
\end{equation}
which is the well-known CHSH inequality \citep{Clauser1969,Nielsen2011}.
It can be written in matrix form as
\begin{equation}
\left|\left\langle \left[\begin{array}{c}
a_{0}\\
a_{1}
\end{array}\right]^{\dagger}\left[\begin{array}{cr}
1 & 1\\
1 & -1
\end{array}\right]\left[\begin{array}{c}
b_{0}\\
b_{1}
\end{array}\right]\right\rangle \right|\le2.\label{eq:CHSHmatrix}
\end{equation}

Quantizing our model, suppose that our bipartite system is made up
two spin-$\frac{1}{2}$ particles, shared between Alice and Bob, with
measurement results governed by quantum theory. Let $a_{0},a_{1}$
represent the spin of Alice's particle along directions of unit vectors
$\hat{a}_{0},\hat{a}_{1}$, and $b_{0},b_{1}$ represent the spin
of Bob's particle along $\hat{b}_{0},\hat{b}_{1}$. Alice (Bob) choose
the spin measurement direction $\hat{a}_{0}$ or $\hat{a}_{1}$ ($\hat{b}_{0}$
or $\hat{b}_{1}$). Assuming Alice and Bob share a singlet state $|\psi\rangle=\frac{1}{\sqrt{2}}\big(|\uparrow\downarrow\rangle-|\downarrow\uparrow\rangle\big)$,
expectation values of joint measurements are %
\begin{comment}
The density matrix of this state may be written as $\rho_{\psi}=\big(I\otimes I-\sigma_{1}\otimes\sigma_{1}-\sigma_{2}\otimes\sigma_{2}-\sigma_{3}\otimes\sigma_{3}\big)/4.$
Then the expectation value of the spin along arbitrary vectors $\vec{c}$
and $\vec{d}$ as measured with a shared singlet state is given by
$\Tr\big[(\vec{c}\cdot\vec{\sigma})\otimes(\vec{d}\cdot\vec{\sigma})\rho_{\psi}\big].$
Simplifying this through the Pauli matrix trace identities $\Tr\sigma_{m}=0$,
and $\Tr\sigma_{m}\sigma_{n}=2\delta_{mn}$, with $\delta_{mn}$ the
Kronecker delta, the expectation value of spins of the singlet state
may be compactly summarized
\end{comment}
{} 
\begin{equation}
\langle(\vec{a}\cdot\vec{\sigma})\otimes(\vec{b}\cdot\vec{\sigma})\rangle_{\psi}=-\vec{a}\cdot\vec{b},\label{eq:singletExp}
\end{equation}
for any $\vec{a}$ and $\vec{b}$. Choosing 
\begin{equation}
\hat{a}_{0}=\left(\begin{smallmatrix}1\\
0\\
0
\end{smallmatrix}\right),\:\hat{a}_{1}=\left(\begin{smallmatrix}0\\
1\\
0
\end{smallmatrix}\right),\:\hat{b}_{0}=\tfrac{1}{\sqrt{2}}\left(\begin{smallmatrix}1\\
1\\
0
\end{smallmatrix}\right),\:\hat{b}_{1}=\tfrac{1}{\sqrt{2}}\left(\begin{smallmatrix}1\\
-1\\
0
\end{smallmatrix}\right),\label{eq:CHSHVectors}
\end{equation}
and using \prettyref{eq:singletExp} to quantize the expression \prettyref{eq:CHSH},
we have 
\begin{equation}
|\langle L\rangle_{\psi}|=|-\hat{a}_{0}{\cdot}\hat{b}_{0}-\hat{a}_{0}{\cdot}\hat{b}_{1}-\hat{a}_{1}{\cdot}\hat{b}_{0}+\hat{a}_{1}{\cdot}\hat{b}_{1}|=2\sqrt{2}.\label{eq:CHSHviolated}
\end{equation}
Therefore, quantum mechanics violates the classical bound of the CHSH
inequality \prettyref{eq:CHSH} by a factor of $\sqrt{2}$, its maximal
possible violation \citep{Cirelson1980}, famously indicating that
quantum theory is not locally real. 

We now introduce additional choices for Alice and Bob to measure the
spin along directions $\hat{a}_{2}$ and $\hat{b}_{2}$ with results
$a_{2},b_{2}\in\{1,-1\}$ respectively. We call this case second order
and the CHSH inequality first order.

In \prettyref{eq:Lfactored}, we made the simple yet useful observation
that exactly one of two quantities $b_{0}+b_{1}$ and $b_{0}-b_{1}$
was nonzero, and took on value $\pm2$. We seek analogous quantities
that include $b_{2}$. Consider the test expression 
\begin{equation}
b_{0}+b_{1}+b_{2}+b_{0}b_{1}b_{2}.\label{eq:expr}
\end{equation}
It takes the value $\pm4$ if $b_{0}=b_{1}=b_{2}=\pm1$, and is $0$
otherwise. Of the eight ($2^{3}$) possible realizations of the triple
$b_{0},b_{1},b_{2}$, only two of them lead to a nonzero value for
the expression \prettyref{eq:expr}. 

We produce three additional test expressions from \prettyref{eq:expr}
by flipping the sign of $b_{1}$, $b_{2}$ or both. For any realization
of $b_{0},b_{1},b_{2}$, exactly one of the four total test expressions
has a nonzero value, equal to $\pm4$. This is outlined in Table \ref{tab:exprTable}.

\begin{table}[h]
\begin{tabular}{|c|rrrrrrrr|}
\hline 
$b_{0}$ & $+1$ & $+1$ & $+1$ & $+1$ & $-1$ & $-1$ & $-1$ & $-1$\tabularnewline
$b_{1}$ & $+1$ & $+1$ & $-1$ & $-1$ & $+1$ & $+1$ & $-1$ & $-1$\tabularnewline
$b_{2}$ & $+1$ & $-1$ & $+1$ & $-1$ & $+1$ & $-1$ & $+1$ & $-1$\tabularnewline
\hline 
$b_{0}+b_{1}+b_{2}+b_{0}b_{1}b_{2}$ & $+4$ & $0$ & $0$ & $0$ & $0$ & $0$ & $0$ & $-4$\tabularnewline
$b_{0}-b_{1}+b_{2}-b_{0}b_{1}b_{2}$ & $0$ & $0$ & $+4$ & $0$ & $0$ & $-4$ & $0$ & $0$\tabularnewline
$b_{0}+b_{1}-b_{2}-b_{0}b_{1}b_{2}$ & $0$ & $+4$ & $0$ & $0$ & $0$ & $0$ & $-4$ & $0$\tabularnewline
$b_{0}-b_{1}-b_{2}+b_{0}b_{1}b_{2}$ & $0$ & $0$ & $0$ & $+4$ & $-4$ & $0$ & $0$ & $0$\tabularnewline
\hline 
\end{tabular}

\caption{\label{tab:exprTable} Values of the four test expressions for the
eight possible realizations of $b_{0},b_{1},b_{2}$. For each realization,
only one of the test expressions is nonzero, and takes the value $\pm4$.}
\end{table}

We then define the quantity $L_{2}$, as the sum of products of the
four test expressions with the four factors \textbf{$a_{0},a_{1},a_{2},$}
and $a_{0}a_{1}a_{2}$\textbf{ }respectively. Since each factor has
value $\pm1$, and only one of the test expressions has a nonzero
value $\pm4$, we conclude $L_{2}=\pm4$ for any given realization.
Averaging all possible realizations of $a_{m}$ and $b_{n}$ in the
ensemble we have 
\begin{equation}
|\langle L_{2}\rangle|=\left|\left\langle \left[\begin{array}{c}
a_{0}\\
a_{1}\\
a_{2}\\
a_{0}a_{1}a_{2}
\end{array}\right]^{\dagger}\left[\begin{array}{rrrr}
1 & 1 & 1 & 1\\
1 & -1 & 1 & -1\\
1 & 1 & -1 & -1\\
1 & -1 & -1 & 1
\end{array}\right]\left[\begin{array}{c}
b_{0}\\
b_{1}\\
b_{2}\\
b_{0}b_{1}b_{2}
\end{array}\right]\right\rangle \right|\le4.\label{eq:O2matrix}
\end{equation}

This matrix inequality is the sought-after generalization of \prettyref{eq:CHSHmatrix},
and must hold under assumptions of local realism. Note that the entries
of the matrix in \prettyref{eq:O2matrix} are the coefficients of
the individual terms in the lower part of the left column in Table
\ref{tab:exprTable}. 

\begin{comment}
Rough:

In practice noncommuting product some ambiguity, rules of correspondence,
Margenau Cohen shows different rules used \citep{Margenau1961,Margenau1967},
applies symmetrization rule in Margenau Hill ... we extend it to product
of three or more ... require symmetric identity in terms of sums and
powers, and linearity of quantization operation, and that quantization
of an (integer) power is the power of the quantization.

Some inconsistency in quantization for continuous variables, \citep{Kapuscik1988,GarciaAlvarez1990}.
but symmetrization makes most sense, our discrete variables makes
it easier (no power of same observable). replace nonlocal with nonclassical
in the manuscript .. and local with classical.

Goal is to find an expression whose violation under qm much higher
than under classical hidden variables. because locality, noncommutativity 
\end{comment}

We now quantize this expression by applying the symmetrization heuristic.
Defining $\vec{a}_{012},\vec{b}_{012}$ as in \prettyref{eq:a012Def},
and making use of \prettyref{eq:singletExp} to evaluate the quantization
of the expression in \prettyref{eq:O2matrix}, we have 
\begin{equation}
|\langle L_{2}\rangle_{\psi}|=\left|\left[\begin{array}{c}
\hat{a}_{0}\\
\hat{a}_{1}\\
\hat{a}_{2}\\
\vec{a}_{012}
\end{array}\right]^{\dagger}\left[\begin{array}{rrrr}
1 & 1 & 1 & 1\\
1 & -1 & 1 & -1\\
1 & 1 & -1 & -1\\
1 & -1 & -1 & 1
\end{array}\right]\left[\begin{array}{c}
\hat{b}_{0}\\
\hat{b}_{1}\\
\hat{b}_{2}\\
\vec{b}_{012}
\end{array}\right]\right|,\label{eq:O2matrixQ}
\end{equation}
where the transpose also applies to the unit vectors inside the left
supervector, and $\hat{a}_{m}^{\dagger}\hat{b}_{n}=\hat{a}_{m}\cdot\hat{b}_{n}$. 

Next, assign the following unit vectors
\begin{equation}
\hat{a}_{0}=\left(\begin{smallmatrix}1\\
0\\
0
\end{smallmatrix}\right),\quad\hat{a}_{1}=\tfrac{1}{2}\left(\begin{smallmatrix}1\\
\sqrt{3}\\
0
\end{smallmatrix}\right),\quad\hat{a}_{2}=\tfrac{1}{2}\left(\begin{smallmatrix}1\\
-\sqrt{3}\\
0
\end{smallmatrix}\right),\quad\hat{b}_{0}=\left(\begin{smallmatrix}-1\\
0\\
0
\end{smallmatrix}\right),\quad\hat{b}_{1}=\tfrac{1}{2}\left(\begin{smallmatrix}-1\\
\sqrt{3}\\
0
\end{smallmatrix}\right),\quad\hat{b}_{2}=\tfrac{1}{2}\left(\begin{smallmatrix}-1\\
-\sqrt{3}\\
0
\end{smallmatrix}\right).\label{eq:O2Vectors}
\end{equation}
These are six maximally separated vectors in the plane, with angle
$\frac{\pi}{3}$ between adjacent vectors. The inner product any two
vectors is the cosine of a multiple of this angle, i.e. $\frac{1}{2}$,
$-\frac{1}{2}$, or $-1$. Indeed $\hat{a}_{0}\cdot\hat{a}_{1}=\hat{a}_{0}\cdot\hat{a}_{2}=\hat{b}_{0}\cdot\hat{b}_{1}=\hat{b}_{0}\cdot\hat{b}_{2}=\tfrac{1}{2},\,\hat{a}_{1}\cdot\hat{a}_{2}=\hat{b}_{1}\cdot\hat{b}_{2}=-\tfrac{1}{2}$,
and the inner product matrix between the $a$ and $b$ vectors is
\begin{equation}
\left[\begin{array}{c}
\hat{a}_{0}^{\dagger}\\
\hat{a}_{1}^{\dagger}\\
\hat{a}_{2}^{\dagger}
\end{array}\right]\left[\begin{array}{ccc}
\hat{b}_{0} & \hat{b}_{1} & \hat{b}_{2}\end{array}\right]=\left[\begin{array}{ccc}
-1 & -\tfrac{1}{2} & -\tfrac{1}{2}\\
-\tfrac{1}{2} & \hphantom{-}\tfrac{1}{2} & -1\\
-\tfrac{1}{2} & -1 & \hphantom{-}\tfrac{1}{2}
\end{array}\right].\label{eq:innerProdMatrix}
\end{equation}
Moreover, for this assignment \prettyref{eq:a012Def} implies $\vec{a}_{012}=\frac{1}{6}(-\hat{a}_{0}+\hat{a}_{1}+\hat{a}_{2})=0.$
Similarly, $\vec{b}_{012}=0$. Therefore, evaluating \prettyref{eq:O2matrixQ}
yields 
\begin{equation}
|\langle L_{2}\rangle_{\psi}|=6.\label{eq:O2is6}
\end{equation}

The quantized result violates the classical bound of $4$ in \prettyref{eq:O2matrix}
by a factor of $\frac{3}{2}$, surpassing the CHSH violation factor
of $\sqrt{2}$. Quantum mechanical violation of classical bounds can
be increased when observables on the same subsystem (qubit) are multiplied,
and their noncommutation is exploited. Put differently,\emph{ }noncommutativity
of quantum operations contributes to quantum violation of classical
realism. Indeed, it has been argued that the product of noncommuting
quantum observables in an isolated system can violate classical realism
without any entanglement \citep{Fine1982,Malley2004,Malley2005}. 

In a similar vein to Tsirelson's bound, Lagrange multipliers show
that the vectors in \prettyref{eq:O2Vectors} are a local maximum
of \prettyref{eq:O2matrixQ}. Numerical optimization suggests it is
a global maximum, and $|\langle L_{2}\rangle_{\psi}|$ cannot exceed
$6$. 

Interestingly, the quantum to classical violation of factor of $\frac{3}{2}$
is precisely the maximal quantum violation of Bell's original inequality
\citep{Bell1964}, which is realized for the same measurement directions
$\hat{a}_{0}$, $\hat{a}_{1}$ and $\hat{a}_{2}$ in \prettyref{eq:O2Vectors}
\citep{Pitowsky2008,Bertlmann2014}. This is despite the very different
forms of \prettyref{eq:O2matrix} and Bell's original inequality,
and the presence of noncommuting products in one but not the other.
\begin{comment}
This raises the question whether $\frac{3}{2}$ is a fundamental quantum
limit or if an inequality can be constructed yielding higher violation? 
\end{comment}

\section{\label{sec:Bell-InequalitiesN}Quasi-Bell inequalities: N+1 Observables}

In search of a higher quantum to classical violation, we generalize
our inequality \prettyref{eq:O2Vectors} to arbitrary order. Let the
$N^{\text{th}}$ order inequality be constructed as follows. Suppose
Alice and Bob share a bipartite system of two spin-$\frac{1}{2}$
particles in a singlet state. They each have $N+1$ measurement options,
yielding results $a_{0},a_{1},\ldots a_{N}\in\{1,-1\}$ and $b_{0},b_{1},\ldots b_{N}\in\{1,-1\}$
respectively. The unit vectors indicating the direction of the quantized
spin operator for $a_{m}$, $b_{n}$ are $\hat{a}_{m}$, $\hat{b}_{n}$. 

Define the scaled Hadamard matrices $M$ and $M_{n}$ as
\begin{equation}
M\equiv\frac{1}{2}\left[\begin{array}{cc}
1 & \hphantom{-}1\\
1 & -1
\end{array}\right],\quad M_{n}\equiv M^{\otimes n},\label{eq:MDef}
\end{equation}
where we used the tensor power. The matrix $M_{n}$ has dimensionality
$2^{n}\times2^{n}$, and satisfies $M_{n}=M\otimes M_{n-1}$. For
completeness, define the trivial matrix $M_{0}\equiv[1]$.

Recursively define the vectors $\vec{A}_{n}$ and $\vec{B}_{n}$ as
\begin{align}
\vec{A}_{n} & \equiv\left[\begin{array}{c}
1\\
a_{0}a_{n}
\end{array}\right]\otimes\vec{A}_{n-1}, & \vec{B}_{n} & \equiv\left[\begin{array}{c}
1\\
b_{0}b_{n}
\end{array}\right]\otimes\vec{B}_{n-1},\label{eq:ABDef}
\end{align}
with $\vec{A}_{0}\equiv[a_{0}],$ $\vec{B}_{0}\equiv[b_{0}]$. Both
$\vec{A}_{n}$ and $\vec{B}_{n}$ are of length $2^{n}$. In the recursion,
we use $a_{n}^{2},b_{n}^{2}=1$ to simplify, leading to no terms having
a power greater than unity. Table \ref{tab:ATable} lists the first
few $\vec{A}_{n}$.

\begin{table}
\begin{tabular}[b]{cccc}
$\vec{A}_{1}$ & $\vec{A}_{2}$ & $\vec{A}_{3}$ & $\vec{A}_{4}$\tabularnewline
\hline 
$\left[\begin{array}{c}
a_{0}\\
a_{1}
\end{array}\right]$ & $\left[\begin{array}{c}
a_{0}\\
a_{1}\\
a_{2}\\
a_{0}a_{1}a_{2}
\end{array}\right]$ & $\left[\begin{array}{c}
a_{0}\\
a_{1}\\
a_{2}\\
a_{0}a_{1}a_{2}\\
a_{3}\\
a_{0}a_{1}a_{3}\\
a_{0}a_{2}a_{3}\\
a_{1}a_{2}a_{3}
\end{array}\right]$ & $\left[\begin{array}{c}
\vec{A}_{3}\\
a_{4}\\
a_{0}a_{1}a_{4}\\
a_{0}a_{2}a_{4}\\
a_{1}a_{2}a_{4}\\
a_{0}a_{3}a_{4}\\
a_{1}a_{3}a_{4}\\
a_{2}a_{3}a_{4}\\
a_{0}a_{1}a_{2}a_{3}a_{4}
\end{array}\right]$\tabularnewline
\end{tabular}\caption{\label{tab:ATable}The vector $\vec{A}_{n}$ as defined in \prettyref{eq:ABDef}
for the first few $n$.}
\end{table}

\begin{comment}
\begin{table}
\begin{tabular}{l|l}
$\vec{A}_{1}$ & $\left[\begin{array}{cc}
a_{0} & a_{1}\end{array}\right]^{\dagger}$\tabularnewline
$\vec{A}_{2}$ & $\left[\begin{array}{cccc}
a_{0} & a_{1} & a_{2} & a_{0}a_{1}a_{2}\end{array}\right]^{\dagger}$\tabularnewline
$\vec{A}_{3}$ & $\left[\begin{array}{cccccccc}
a_{0} & a_{1} & a_{2} & a_{0}a_{1}a_{2} & a_{3} & a_{0}a_{1}a_{3} & a_{0}a_{2}a_{3} & a_{1}a_{2}a_{3}\end{array}\right]^{\dagger}$\tabularnewline
$\vec{A}_{4}$ & $\begin{array}{ccccccc}
\Big[\vec{A}_{3} & a_{4} & a_{0}a_{1}a_{4} & a_{0}a_{2}a_{4} & a_{1}a_{2}a_{4} & a_{0}a_{3}a_{4} & \ldots\\
 &  &  & \ldots & a_{1}a_{3}a_{4} & a_{2}a_{3}a_{4} & a_{0}a_{1}a_{2}a_{3}a_{4}\Big]^{\dagger}
\end{array}$\tabularnewline
\end{tabular}\caption{The vector $\vec{A}_{n}$ as defined in \prettyref{eq:ABDef} for
the first few $n$.}
\end{table}
\end{comment}

Define the $N^{\text{th}}$ order quantity $K_{N}\equiv\vec{A}_{N}^{\dagger}M_{N}\vec{B}_{N}$.
Given the above, it can be simplified as follows 
\begin{align}
K_{N} & =\vec{A}_{N}^{\dagger}M_{N}\vec{B}_{N}\nonumber \\
= & \Bigg(\left[\begin{array}{c}
1\\
a_{0}a_{N}
\end{array}\right]^{\dagger}{\otimes}\vec{A}_{N-1}^{\dagger}\Bigg)M{\otimes}M_{N-1}\Bigg(\left[\begin{array}{c}
1\\
b_{0}b_{N}
\end{array}\right]{\otimes}\vec{B}_{N-1}\Bigg)\nonumber \\
= & a_{0}b_{0}\frac{1}{2}\left[\begin{array}{c}
a_{0}\\
a_{N}
\end{array}\right]^{\dagger}\left[\begin{array}{cc}
1 & \hphantom{-}1\\
1 & -1
\end{array}\right]\left[\begin{array}{c}
b_{0}\\
b_{N}
\end{array}\right]\otimes\vec{A}_{N-1}^{\dagger}M_{N-1}\vec{B}_{N-1}\nonumber \\
= & \pm K_{N-1},\label{eq:BellQuantityRecursion}
\end{align}
where in the last line we noted $a_{0}b_{0}=\pm1$, and the matrix
multiplication on the left side of tensor product is a CHSH term taking
values $\pm2$. Extending the recursion,
\[
K_{N}=\pm K_{N-1}=\ldots=\pm K_{0}=\pm a_{0}b_{0}=\pm1.
\]

Since for each realization of the $a_{n}$ and $b_{n}$, $K_{N}$
is $\pm1$, the average over the classical ensemble satisfies
\begin{equation}
|\langle K_{N}\rangle|=|\langle\vec{A}_{N}^{\dagger}M_{N}\vec{B}_{N}\rangle|\le1.\label{eq:ONinequality}
\end{equation}
Comparing \prettyref{eq:ONinequality} with its precursors \prettyref{eq:O2matrixQ}
and \prettyref{eq:CHSHmatrix}, we see a power of $2$ scaling difference.
The factor of $\frac{1}{2}$ in the definition of $M$ \prettyref{eq:MDef}
cancels this power, and ensures the classical expectation value has
absolute value at most unity for all orders. This facilitates comparison
of inequalities of different orders. 

Finally, we quantize the expression for $K_{N}$, and maximize its
quantum expectation value over all possible measurement choices $\hat{a}_{n}$,
$\hat{b}_{n}$. Products of noncommuting observables are always quantized
through symmetrization, analogous to \prettyref{eq:a012Def}. We obtained
numerical results up to $N=10$, shown in Table \ref{tab:ONOptimized}. 

The $0^{\text{th}}$ order is the classical case, the $1^{\text{st}}$
order violation is the CHSH value of $\sqrt{2}$, and $2^{\text{nd}}$
order yields $\frac{3}{2}$ violation demonstrated above. Interestingly,
optimized violation ratios for higher orders always lie between the
$1^{\text{st}}$ and $2^{\text{nd}}$ order cases.

\begin{table}[h]
\begin{tabular}{c|ccccccccccc}
Order $N$ & $\;0\;$ & $1$ & $\:2\:$ & $3$ & $4$ & $5$ & $6$ & $7$ & $8$ & $9$ & $10$\tabularnewline
\hline 
$\max|\langle K_{N}\rangle_{\psi}|$ & $1$ & $1.414$ & $\mathbf{1.5}$ & $1.432$ & $1.469$ & $1.443$ & $1.467$ & $1.45$ & $1.467$ & $1.455$ & $1.469$\tabularnewline
\noalign{\vskip0.2cm}
\end{tabular}

\caption{\label{tab:ONOptimized}The maximized quantum expectation value $|\langle K_{N}\rangle_{\psi}|$
to three decimal places for $N$ up to $10$.}
\end{table}

The optimal measurement vectors in the $1^{\text{st}}$ and $2^{\text{nd}}$
order cases, in \prettyref{eq:CHSHVectors} and \prettyref{eq:O2Vectors}
respectively, were coplanar. The optimal vectors for higher orders
also turn out to be coplanar. More precisely, we found that the optimal
vectors satisfy $\hat{a}_{2}=\hat{a}_{3}=\ldots=\hat{a}_{N},\,\hat{b}_{2}=\hat{b}_{3}=\ldots=\hat{b}_{N}.$
Thus, for each qubit subsystem, the $N+1$ optimized spin measurement
vectors include only three unique (and coplanar) vectors, as $N-1$
of them are identical. This gives insight into why going beyond $2^{\text{nd}}$
order actually decreases the violation ratio; adding more measurement
options beyond three only replicates an existing measurement option
when optimized. Further, product vectors $\vec{a}_{012}$, $\vec{a}_{013}$,
$\vec{a}_{123}$ etc. cannot all simultaneously be optimized to zero
as in the $2^{\text{nd}}$ order case. 

If this family of inequalities turns out to yield the highest possible
violation, it would imply that adding a third spatial dimension to
a two dimensional system does not increase the maximal possible ``non-classicality''.
It is unclear what the effect of additional spatial dimensions beyond
three would be.

\section{\label{sec:Classicality-of-Werner}Locality of Werner States}

Here we put the findings in perspective by comparing them to existing
work on ranges of locality. We start by adding noise to our singlet
state $|\psi\rangle$, by replacing it with the two qubit Werner state
\[
\rho(z)=\frac{1-z}{4}I+z|\psi\rangle\langle\psi|.
\]
This results in the expectation value \prettyref{eq:singletExp} changing
to 
\begin{equation}
\langle(\vec{c}\cdot\vec{\sigma})\otimes(\vec{d}\cdot\vec{\sigma})\rangle_{\rho(z)}=-z\vec{c}\cdot\vec{d}.\label{eq:WernerExp}
\end{equation}
 Therefore all quantum expectation values are now scaled by a factor
of $z$. 

A rich research program has followed from investigating the lowest
value of $z$ for which the Werner state violates a Bell inequality,
and the highest value for which a local hidden variable model can
be constructed. With the progress of research, the two values have
been converging, but have yet to meet.

The CHSH inequality implies a nonclassicality range of $z>\frac{1}{\sqrt{2}}\approx0.7071$,
V\'ertesi's slightly improved the range to $z>0.7056$ \citep{Vertesi2008},
and Brierley et al. to $z>0.7012$ \citep{Brierley2016}. It is worth
noting that although John Bell's original inequality matches our $\frac{3}{2}$
violation for a singlet state \citep{Bell1964}, it loses much of
its efficacy for Werner states, yielding a weaker nonclassicality
range for $z$ than the CHSH inequality \citep{Pitowsky2008}.

On the other end of the spectrum, the Werner state is separable, and
therefore local, for $z\le\frac{1}{3}$. By explicit construction
of a simple hidden variable model, Werner showed that it is local
for $z\le\frac{1}{2}$ \citep{Werner1989}, with the surprising implication
that an entangled state could have a local hidden variable description.
Toner et al. constructed an even stronger hidden variable model demonstrating
locality for $z\lesssim0.6595$ \citep{TonerLHV2005,Acin2006}. Most
recently, Hirsch et al. recently reported locality for $z\lesssim0.6829$
\citep{Hirsch2016,Hirsch2017}, meaning the boundary between the local
and nonlocal Werner states must lie between $0.6829$ and $0.7012$.

\begin{comment}
Figure \ref{fig:zSpectrum} summarizes the $z$ spectrum.

Attempt to draw it using tikzpackage

\textbackslash{}begin\{tikzpicture\}{[}scale=3{]} \textbackslash{}draw
{[}thin, gray, -latex{]} (0,0) \textendash{} (1.1,0); \textbackslash{}foreach
\textbackslash{}Tick in \{0,\$\textbackslash{}frac\{1\}\{3\}\$,\$\textbackslash{}frac\{1\}\{2\}\$,0.6596,2/3,1\}
\{ \textbackslash{}draw {[}tick style{]} (\textbackslash{}Tick,1.5ex)
\textendash{} (\textbackslash{}Tick,-1.5ex) node {[}below{]} \{\$\textbackslash{}Tick\$\}
; \} \textbackslash{}end\{tikzpicture\}
\end{comment}

It is now natural to ask where our result lies along this spectrum.
The expectation value \prettyref{eq:WernerExp} changes our maximal
quantum to classical violation to $\frac{3}{2}z$. This means the
Werner state violates our Quasi-Bell inequality for $z>\frac{2}{3}$.
It may then seem that our result contradicts the range of locality
given by Hirsch et al. But, in fact, this is to be expected; the additional
burden of producing joint probabilities consistent with quantum theory
weakens our local hidden variable models relative to standard ones,
leading to a smaller range of locality.

The above local hidden variable models by various authors do not define
joint probability distributions for noncommuting observables as we
do. Therefore, they should not be compared at face value with our
$z>\frac{2}{3}$ nonlocality range. 

Nonetheless, addressing joint probabilities is a burden local hidden
variable models should be able to bear. It is in principle possible
to extend standard local hidden variable models to do this. Defining
joint probabilities (e.g. $p(a_{0},a_{1}|\hat{a}_{0},\hat{a}_{1})$)
that imply the marginal probabilities ($p(a_{0}|\hat{a}_{0}),$$p(a_{1}|\hat{a}_{1})$)
in the existing models would achieve this. The joint probabilities
would be constructed to respect some desired criteria, such as positivity,
the Fr\'echet inequalities, or, more likely, the symmetrization procedure,
in which case the resulting local hidden variable model must satisfy
our Quasi-Bell inequalities. 

\begin{comment}
Therefore, the critical value $z_{c}$ for which the Werner state
is local for $z\le z_{c}$ and nonlocal for $z>z_{c}$ must lie within
the narrow range $0.6595\lesssim z_{c}\le\frac{2}{3}$. 

Newton proclaimed that nature is pleased with simplicity. The upper
bound we just placed on $z_{c}$ is simpler in value and derivation
than the lower. One is tempted to take Newton's adage to heart, and
conjecture that $z_{c}=\frac{2}{3}$. However, the value of $z_{c}$
will only be known with certainty if we find a tighter Bell inequality
or a stronger hidden variable model to close the gap. 
\end{comment}

\section{\label{sec:Discussion}Discussion }

Here we summarize our results, discussing their meaning and significance.
We began by reviewing the concept of joint probabilities for commuting
observables, and relating the quantum probabilities with those due
to classical local hidden variables. We then examined joint probabilities
of noncommuting observables, showing that in general they require
some assumptions to determine them from the marginal probabilities. 

We found that assuming the joint probability of noncommuting observables
obeys the Fr\'echet inequalities is equivalent to assuming its positivity.
It turns out that for pure single-qubit states positivity implies
independence of measurement results from noncommuting observables.
Since independence is problematic, we sought other avenues to calculating
the joint probability. This resulted in negative joint probabilities,
well known for noncommuting observables, which do not lead to physical
contradictions as they cannot be directly measured.

We introduced symmetrization as a means to quantization of classical
products of noncommuting observables, and justified it heuristically
then more carefully via Moyal quantization. The justification of symmetrization
is interesting in its own right, independent of Bell inequalities.
It may lead to its own research trajectory where one tests symmetrization
through physical Hamiltonians that include products of noncommuting
observables. Although symmetrization mathematically follows from application
of canonical quantization, it remains to be experimentally tested.

Based on symmetrization, we created a hierarchy of Quasi-Bell inequalities,
which yields a quantum to classical violation higher than standard
Bell inequalities. Moreover, the family of Quasi-Bell inequalities
in Sections \ref{sec:Bell-Inequalities-Three} and \ref{sec:Bell-InequalitiesN}
exceeds treatments relying on Grothendieck inequalities \citep{Grothendieck1953,Krivine1979,Tsirelson1987}.
This is because the latter are restricted to unit vectors, while in
this work, the vectors derived from quantization of noncommuting products,
such as $\vec{a}_{012}$, are not generally of unit magnitude. 

Although our Quasi-Bell inequality yields a higher violation, it cannot
be compared with standard Bell inequalities or local hidden variable
models on Werner states, because the latter do not consider joint
probabilities of noncommuting observables.%
\begin{comment}
A celebrated aspect of the CHSH Bell inequality is that it subjects
fundamental questions to experiment \citep{Freedman1972,Aspect1981c,Aspect1982,Aspect1982b,Hensen2015,Shalm2015b}.
The use of noncommuting products in the model presented here means
operationally measuring the symmetrized observables, $\vec{a}_{012}{\cdot}\vec{\sigma}$
and $\vec{b}_{012}{\cdot}\vec{\sigma}$, which reduce to zero for
the optimal settings. 

It is crucial to note that the validity of the conclusions from our
Quasi-Bell inequalities rests on the heuristic symmetrization of \prettyref{sec:Symmetrization}.
Symmetrization is consistent with the axioms of quantum theory, and
almost universally applied for quantization of products of noncommuting
classical observables. Extending symmetrization to hidden variable
models is the most reasonable and straightforward assumption when
dealing with noncommuting products, but not the only one that can
be made. Therefore, despite the higher violation, the conceptual conclusions
from our results are not as strong as those from conventional Bell
inequalities
\end{comment}
{} 

Our findings shed important light on joint distributions of noncommuting
observables and their consequent extensions of Bell inequalities.
The two essential ``strange'' features of quantum theory, noncommutativity
and nonlocality, are shown to influence one another in interesting
ways. The findings also help us better understand the limitations
of local hidden variable models, and the possibility of extending
them to include joint probabilities. We showed interesting examples
of negative probabilities that appear in intermediate theoretical
quantities (e.g. joint probability) but not in observable measurements. 

There of course remains the important question, can Quasi-Bell inequalities
be tested experimentally? Traditional Bell inequalities proved that
quantum mechanics, as a theory, is nonlocal. The question then became
whether reality is nonlocal, i.e. is quantum mechanics an accurate
representation of reality? This led to numerous experimental tests
spanning six decades \citep{Freedman1972,Aspect1981c,Aspect1982,Aspect1982b,Hensen2015,Shalm2015b,Giustina2015}
which confirmed the validity of quantum theory. 

However, while the insights of Quasi-Bell inequalities are many, they
remain, thus far, entirely on the theoretical side. Once one decides
to do the experiment to measure the quantities in the Quasi-Bell inequality,
one has to measure symmetrized operators like $\vec{a}_{012}{\cdot}\vec{\sigma}$
and $\vec{b}_{012}{\cdot}\vec{\sigma}$ (which are incidentally set
to zero for the optimal setting for $N=3$), because one cannot simultaneously
measure noncommuting observables to calculate their product. But in
deriving the symmetrization from the noncommuting product, we used
canonical quantization, thereby \emph{assuming the validity of quantum
theory,} and as a consequence, nonlocality. Therefore such an experiment
would be implicitly assuming nonlocality and cannot be used to test
it.

This will remain true unless%
\begin{comment}
simultaneous 
\end{comment}
{} direct measurement of the product of non commuting observables %
\begin{comment}
somehow 
\end{comment}
{} becomes experimentally%
\begin{comment}
 (and theoretically)
\end{comment}
{} possible. In which case direct measurement of the non commuting product
will circumvent the need for symmetrization. Thankfully, this is not
a serious problem; great strides by experimentalists confirming the
violation of traditional Bell inequalities mean the nonlocality of
nature is no longer in doubt. 
\begin{acknowledgments}
We thank Toner et al. for sharing their unpublished manuscript \citep{TonerLHV2005}.
We also thank the anonymous referees and Dr. Michael Hall for insightful
critiques. This work was supported by the Director, Office of Science,
Office of Basic Energy Sciences, of the USA Department of Energy under
Contract No. DE-AC02-05CH11231 and the Division of Chemical Sciences,
Geosciences and Biosciences Division, Office of Basic Energy Sciences
through Grant No. DE-AC03-76F000098 (at LBNL and UC Berkeley).
\end{acknowledgments}

\appendix

\section*{Moyal Quantization of the Product }

In this appendix, we simplify the expression in \prettyref{eq:MomentQuantum}
to find the operator corresponding to Moyal quantization of the product
$a_{0}a_{1}a_{2}$.

We define 
\begin{align}
\vec{\chi} & \equiv\theta_{0}\hat{a}_{0}+\theta_{1}\hat{a}_{1}+\theta_{2}\hat{a}_{2},\nonumber \\
\chi\equiv|\vec{\chi}| & =\sqrt{\theta_{0}^{2}{+}\theta_{1}^{2}{+}\theta_{2}^{2}{+}2(\hat{a}_{0}{\cdot}\hat{a}_{1})\theta_{0}\theta_{1}{+}2(\hat{a}_{0}{\cdot}\hat{a}_{2})\theta_{0}\theta_{2}{+}2(\hat{a}_{1}{\cdot}\hat{a}_{2})\theta_{1}\theta_{2}}.\label{eq:MoyalVecMagnitude}
\end{align}
Then using the well-known identity for exponents of Pauli vectors,
our quantum operator in \prettyref{eq:MomentQuantum} becomes
\begin{equation}
\left[\frac{\partial^{3}}{i^{3}\partial\theta_{0}\partial\theta_{1}\partial\theta_{2}}\left(\cos\chi\,I+i\,\frac{\sin\chi}{\chi}\,(\vec{\chi}\cdot\vec{\sigma})\right)\right]^{\theta_{i}{=}0}.\label{eq:MoyalOperator1}
\end{equation}
Considering only the cosine term, we expand its Taylor series to get
\begin{equation}
\left[\frac{\partial^{3}}{i^{3}\partial\theta_{0}\partial\theta_{1}\partial\theta_{2}}\left(1-\frac{1}{2!}(\theta_{0}^{2}{+}\theta_{1}^{2}{+}\theta_{2}^{2}{+}2(\hat{a}_{0}{\cdot}\hat{a}_{1})\theta_{0}\theta_{1}{+}2(\hat{a}_{0}{\cdot}\hat{a}_{2})\theta_{0}\theta_{2}{+}2(\hat{a}_{1}{\cdot}\hat{a}_{2})\theta_{1}\theta_{2})+O(\theta^{4})\right)\right]^{\theta_{i}{=}0}I,\label{eq:MoyalCosine}
\end{equation}
where $O(\theta^{4})$ denotes terms with products of four or more
$\theta_{i}$ . Since the Taylor expansion lacks a $\theta_{0}\theta_{1}\theta_{2}$
term, it is clear that the expression \prettyref{eq:MoyalCosine}
vanishes. We are then left with the more interesting sine term, which
we also expand in a Taylor series and simplify as

\begin{align}
 & \left[\frac{\partial^{3}}{i^{2}\partial\theta_{0}\partial\theta_{1}\partial\theta_{2}}\left(1-\frac{1}{3!}(\theta_{0}^{2}{+}\theta_{1}^{2}{+}\theta_{2}^{2}{+}2(\hat{a}_{0}{\cdot}\hat{a}_{1})\theta_{0}\theta_{1}{+}2(\hat{a}_{0}{\cdot}\hat{a}_{2})\theta_{0}\theta_{2}{+}2(\hat{a}_{1}{\cdot}\hat{a}_{2})\theta_{1}\theta_{2})+O(\theta^{4})\right)\left(\theta_{0}\hat{a}_{0}+\theta_{1}\hat{a}_{1}+\theta_{2}\hat{a}_{2}\right)\right]^{\theta_{i}{=}0}\cdot\vec{\sigma}\nonumber \\
 & =\frac{1}{3}\left[\frac{\partial^{3}}{\partial\theta_{0}\partial\theta_{1}\partial\theta_{2}}\left((\hat{a}_{0}{\cdot}\hat{a}_{1})\theta_{0}\theta_{1}{+}(\hat{a}_{0}{\cdot}\hat{a}_{2})\theta_{0}\theta_{2}{+}(\hat{a}_{1}{\cdot}\hat{a}_{2})\theta_{1}\theta_{2}\right)\left(\theta_{0}\hat{a}_{0}+\theta_{1}\hat{a}_{1}+\theta_{2}\hat{a}_{2}\right)\right]^{\theta_{i}{=}0}\cdot\vec{\sigma}\nonumber \\
 & =\frac{1}{3}\left[(\hat{a}_{1}{\cdot}\hat{a}_{2})\hat{a}_{0}+(\hat{a}_{2}{\cdot}\hat{a}_{0})\hat{a}_{1}+(\hat{a}_{0}{\cdot}\hat{a}_{1})\hat{a}_{2}\right]\cdot\vec{\sigma}\nonumber \\
 & =\vec{a}_{012}\cdot\vec{\sigma},\label{eq:MoyalSine}
\end{align}
where in going to the second line we noted that $\theta_{i}^{2}$
and $O(\theta^{4})$ terms cannot contribute to the $\theta_{0}\theta_{1}\theta_{2}$
term, and the final line is exactly that defined in \prettyref{eq:symmetrization-three}.

\end{document}